\documentclass[%
 reprint,
 amsmath,amssymb,
 aps,
]{revtex4-2}

\usepackage{graphicx}
\usepackage{dcolumn}
\usepackage{bm}

\begin{document}

\preprint{APS/123-QED}

\title{Bjorken initial energy density estimation in Xe-Xe collisions at $\sqrt{s_{NN}} = 5.44$ TeV using ALICE}

\author{Mohammad Asif Bhat$^{1}$}
\email{asifqadir1994@gmail.com; mohammad.bhat@iopb.res.in}
\author{Akankshya Nayak$^{1,2}$}
\author{P. K. Sahu$^{1}$}%
\email{pradip@iopb.res.in}
\affiliation{${}^1$Institute of Physics, Sachivalaya Marg, Sainik School P.O., Bhubaneswar 751005, India\\
Homi Bhaba National Institute Anushakti Nagar, Mumbai, 400094, India.\\ 
${}^2$Department of Physics, School of Applied Sciences, KIIT, Deemed to be University, Bhubaneswar 751024, India.}




\date{\today}

\begin{abstract}
The Bjorken initial energy density ($\epsilon_{B}$) has been estimated in different centrality classes in Xe-Xe collisions using Bjorken formula. Three different cases have been considered. In Case I, we have fixed the formation time ($\tau_{0}$) and varied the area of overlap region ($A_{overlap}$). In Case II, we have fixed both $\tau_{0}$ and $A_{overlap}$ and we have varied both $\tau_{0}$ and $A_{overlap}$ in Case III. We observed that the $\epsilon_{B}$ is first increasing and then decreasing while going from central to peripheral collisions in Case I whereas it is decreasing in Case II and III, as expected. The $\epsilon_{B}$ value for top central and mid central collisions in Case II and for all centrality classes in Case III are greater than 1 $GeV/(fm)^{3}$, indicating the possibility of QGP medium formation respectively. The $\epsilon_{B}$ results of the Pb-Pb at $\sqrt{s_{NN}} = 2.76$ and 5.02 TeV are compared. In Case II, it is observed that in the centrality class (0-5)\%, the $\epsilon_{B}$ value in Xe-Xe collisions at $\sqrt{s_{NN}} = 5.44$ TeV is 34.66\% less than in Pb-Pb collisions at $\sqrt{s_{NN}} = 2.76$ TeV and 57.12\% less than at 5.02 TeV. This may be interpreted as, the system size in Xe-Xe collisions at $\sqrt{s_{NN}} = 5.44$ TeV is 34.66\% smaller than in Pb-Pb collisions at $\sqrt{s_{NN}} = 2.76$ TeV and 57.12\% smaller than at 5.02 TeV. Similarly in Case III in the centrality class (0-5)\%, the $\epsilon_{B}$ value in Xe-Xe collisions at $\sqrt{s_{NN}} = 5.44$ TeV is 3.22\% more than in Pb-Pb collisions at $\sqrt{s_{NN}} = 2.76$ TeV and 32.18\% less than at 5.02 TeV. This may be interpreted as, the system size in Xe-Xe collisions at $\sqrt{s_{NN}} = 5.44$ TeV is 3.22\% bigger than in Pb-Pb collisions at $\sqrt{s_{NN}} = 2.76$ TeV and 32.18\% smaller than at 5.02 TeV. The results obtained in the Case III seems more suitable than Case II.  
\end{abstract}

\maketitle


\section{\label{A}Introduction}
In the early universe shortly after the Big Bang, a new state of matter is believed to have existed called Quark-Gluon Plasma (QGP)~\cite{Riordan:2006df,Gale:2013da,Shuryak:2014zxa,Rafelski:1982pu,Koch:1986ud,Koch:1988nn,Bass:2002jy}. Quarks and gluons which are normally confined within protons and neutrons, move freely in this state in a hot and dense soup. The existence of QGP was first proposed in the 1970s~\cite{Shuryak:1978ij}, an attempt to understand the behavior of high energy collisions between atomic nuclei.\par
The experimental evidence for the QGP creation was provided by the Relativistic Heavy Ion Collider (RHIC) in heavy ion collisions such as Au-Au collisions at $\sqrt{s_{NN}} = 200$ GeV~\cite{PHENIX:2004vcz,Bellwied:2005kq,Heinz:2008tv,Muller:2006ee}. These collisions produced a liquid which is made up of quarks and gluons having the ratio of shear viscosity to entropy density lower than any other known liquid~\cite{Gyulassy:2004zy,Nouicer:2015jrf,STAR:2005gfr,Pal:2010es,Tannenbaum:2006ch,Luzum:2008cw,Song:2007fn}. The Large Hadron Collider Experiment (LHC) later confirmed its existence in
Pb-Pb collisions~\cite{ALICE:2010suc,ATLAS:2010isq,ATLAS:2011ah}. In other collisions systems such as pp, p-Pb at LHC~\cite{Kharzeev:2014pha,CMS:2012qk,ALICE:2016fzo,CMS:2016fnw,Bjorken:2013boa,Sahoo:2019ifs} and $^{3}$He-Au, d-Au at RHIC~\cite{PHENIX:2018lia} similar properties have been observed.\par
The QGP system created is driven by pressure gradients, expands collectively and cools until hadronization. The relativistic hydrodynamics well described the spacetime evolution of the QGP~\cite{Busza:2018rrf,Kuroki:2023ebq,Shi:2024pyz,Karimabadi:2023sif,Jaiswal:2016hex}. The geometric anisotropies in the initial state is turned by the gradient driven expansion into anisotropic flow in the final state and the variations in size in the initial state into radial flow~\cite{Prasad:2022zbr,Bozek:2012fw,Samanta:2023amp}. Azimuthal momentum anisotropy is quantified by the anisotropic flow~\cite{ALICE:2011ab,PHENIX:2011yyh,ATLAS:2012at,CMS:2013wjq,ATLAS:2013xzf,ATLAS:2014ndd,ATLAS:2015qwl,ATLAS:2019peb}, while the radial boost of the system, which influences the average transverse momentum of particles in each event is characterized by radial flow~\cite{Heinz:2013th,Broniowski:2009fm,Bozek:2021zim,Waqas:2023baa}.\par
The reaction dynamics can be better understood by the fluid hydrodynamics. In a collision at non-zero impact parameter, the anisotropy of the low $p_{T}$ particles produced is described by the elliptic flow. This suggests that a collective flow of the particles exist following a hydrodynamical pressure gradient which is due to the initial eccentricity in a collision~\cite{Ollitrault:1992bk,Yadav:2025vtc,Shen:2011eg}. The elliptic flow~\cite{Hirano:2008hy} is successfully described by most of the hydrodynamical simulations which are compatible with an almost ``perfect fluid'' behaviour, i.e. a small ratio of viscosity to entropy~\cite{Huovinen:2006jp,Niemi:2011ix,Demir:2008tr}.\par
\begin{figure}[htb]
	\centering
	\includegraphics[width=1\linewidth]{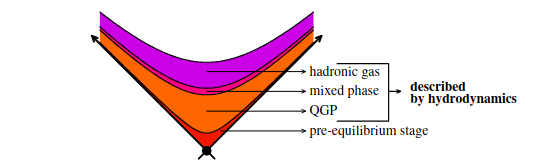}
	\caption{\label{1}QGP medium formation description in heavy ion collisions.~\cite{Heller:2008fg}}
\end{figure}

The hydrodynamical description of the QGP medium created is validated by assuming the quasi-perfect fluid behaviour~\cite{Daher:2024vxk,Capellino:2022nvf}. In a reaction process hypothesis an intermediate stage such as boost invariant QGP phase as a relativistic expanding fluid, is the foundation of the so-called Bjorken flow. After a vert quick thermalization period this QGP phase is formed and finally results into hadrons. The description of QGP medium formation in heavy ion collisions is shown in figure~\ref{1}.
The boost-invariance can be justified in the central region of the collision as the distribution of particles observed is flat. This is consistent with the hydrodynamical prediction of boost-invariance, where rapidities of fluid (space-time) and particle (energy-momentum) are equal.
\section{\label{B}Bjorken Model}
The hydrodynamic description introduced by Bjorken is one of the principal models used to describe the heavy-ion collisions at high energies. The popular Bjorken flow~\cite{Gubser:2012gy,Bagchi:2023ysc,Mitra:2020mei,Gubser:2010ze}, which is the boost invariant fluid flow does not depend on rapidity but only on proper time $(\tau)$. In high energy collisions at central rapidity, this phenomenological assumption holds true. This model has been very successful in describing the extreme regime, where velocity of fluid is close to the light velocity and is one of the simplest models. The spacetime evolution of highly energetic and dense state of matter created in heavy-ion collisions is described by the Bjorken flow as an ultrarelativistic fluid~\cite{Ciambelli:2018xat,Petkou:2022bmz}. All the interesting dynamics takes place along the direction in which the collision of the two heavy nuclei occur i.e. along the beam axis, which is usually taken as $z$-axis. This is considered as one among the major simplifying assumptions of Bjorken flow. In the transverse $x-y$ plane, the flow expects complete rotational and translational invariance, which leads it to become effectively two dimensional. The boost or to be more precise rapidity invariance~\cite{Simeoni:2022hjh}, which claims about the existence of the velocity profile of the produced fluid after the collision, is the second major assumption of the Bjorken flow. The longitudinal velocity of the fluid at the location $z$ is given by $v = \frac{z}{t}$, after assuming that the collision occurred at $z = 0$ and at time $t = 0$. However, at any later time $t$ the fluid exactly in the midway between the two receding nuclei continues to be at rest. This can also be interpreted as, the fluid at $z = 0$ is at rest at a particular instant in time $t$, whereas the fluid is moving with the speed of light at $z = \pm t$. The temperature as well as energy of the fluid decreases with the time as $\tau^{\frac{-1}{3}}$, where $\tau = \sqrt{t^{2} - z^{2}}$ is the proper time. Bjorken considered the QGP medium as fluid system, after applying these assumptions and some other assumptions, he derived the initial energy density formula given as. 
\begin{equation}
	\epsilon_{B} =\; <m_{t}>\frac{3}{2}\frac{dN_{ch}}{dy}\frac{1}{\tau_{o}\pi R^2}\;\;\;\;\;\;\;\;,
\end{equation}
where $<m_{t}>$, is the transverse mass of the produced particles. $\frac{dN_{ch}}{dy}$ is the charged particle rapidity density. $\tau_{0}$ is the formation time of possible hydro type system or initial time of a possible hydro type of evolution. $\pi R^2$ is the area of overlap region~\cite{Bjorken:1982qr}. According to this model, if initial energy density $\epsilon_{B} \ge 1 GeV/(fm)^{3}$ then there exists the QGP medium.\par
The main aim of this work is to investigate the QGP medium formation in Xe-Xe collisions in both central and peripheral collisions using Bjorken limit of 1 $GeV/(fm)^{3}$. Also the comparison of $\epsilon_{B}$ with the Pb-Pb results at $\sqrt{s_{NN}} = 2.76$ TeV~\cite{PHENIX:2015vqa} and $\sqrt{s_{NN}} = 5.02$ TeV~\cite{Prasad:2021bdq} will give the idea of system size formed in Xe-Xe collisions.\par
In this presentation, we have estimated the initial energy density in Xe-Xe collisions at $\sqrt{s_{NN}} = 5.44$ TeV in different centrality classes using the Bjorken formula, for three different cases which is discussed in section~\ref{D}. We have used $<m_{t}>$ as the transverse masss of pions equal to 0.562 GeV. The charged particle rapidity density $\frac{dN_{ch}}{dy}$ used is taken from ALICE~\cite{ALICE:2018cpu}. The formation time $\tau_{0}$ used is 1 $(fm/c)$ for all the centrality classes in Case I and II and for Case III, $\tau_{0}$ used varies from 1 to 2.6 $(fm/c)$, while going from central to the peripheral collisions. The methodology by which radius of the Xe nucleus, radius of overlap region between two colliding nuclei, area of overlap region between two colliding nuclei and Bjorken initial energy density calculated are discussed in section~\ref{C}.
\section{\label{C} Methodology}
\subsection{Radius of Xenon (Xe) nucleus}
The radius of the Xe nucleus is calculated by the nuclear radius equation given as
\begin{equation}
	R = R_{0}A^{\frac{1}{3}}
\end{equation}
Where, R is the nuclear radius, $R_{0}$ is a constant known as the Fermi radius, which is approximately $1.2\times10^{-15}$ meters = 1.02 fm, and A is the nucleon number, which is the total number of neutrons and protons inside the nucleus. The Xenon isotope which was collided in LHC at $\sqrt{s_{NN}} = 5.44$ TeV is $^{129}_{54}Xe$, with number of protons (P) 54 and nucleon number (A) 129.  The number of neutrons (N) = nucleon number (A) - number of protons (P) = 129 - 54 = 75. The radius of Xe nucleus calculated from the above eq. (2) is $6.06\times10^{-15}$ meters = 6.06 fm.
\subsection{Radius of overlap region}
\begin{figure}[htb]
	\centering
	\includegraphics[width=0.62\linewidth]{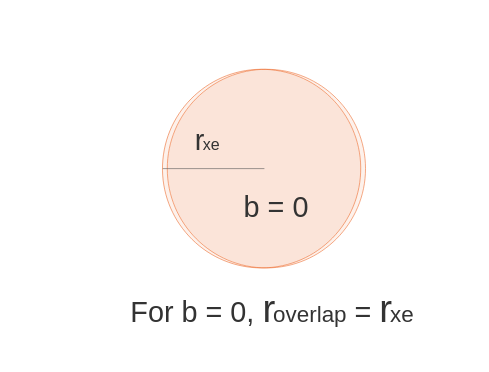}
	\caption{\label{2} Top central collision in which impact parameter (b) = 0 and radius of overlap region ($r_{overlap}) = r_{Xe}$.}
\end{figure}

\begin{figure}[htb]
	\centering
	\includegraphics[width=0.62\linewidth]{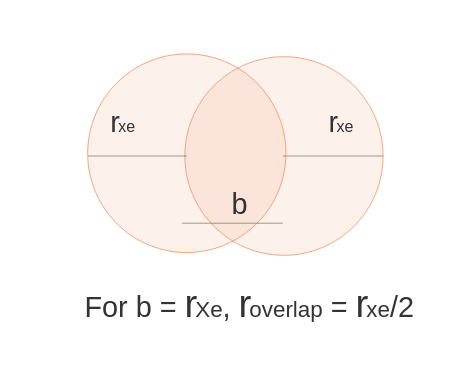}
	\caption{\label{3} Mid central collision in which impact parameter $(b) = r_{Xe}$ and radius of overlap region ($r_{overlap}) = \frac{r_{Xe}}{2}$.}
\end{figure}

\begin{figure}[htb]
	\centering
	\includegraphics[width=0.62\linewidth]{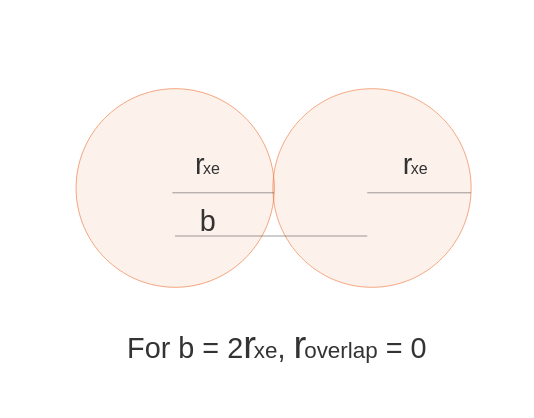}
	\caption{\label{4} Peripheral collision in which impact parameter $(b) = 2r_{Xe}$ and radius of overlap region ($r_{overlap}) = 0$.}
\end{figure}

\begin{figure}[htb]
	\centering
	\includegraphics[width=0.91\linewidth]{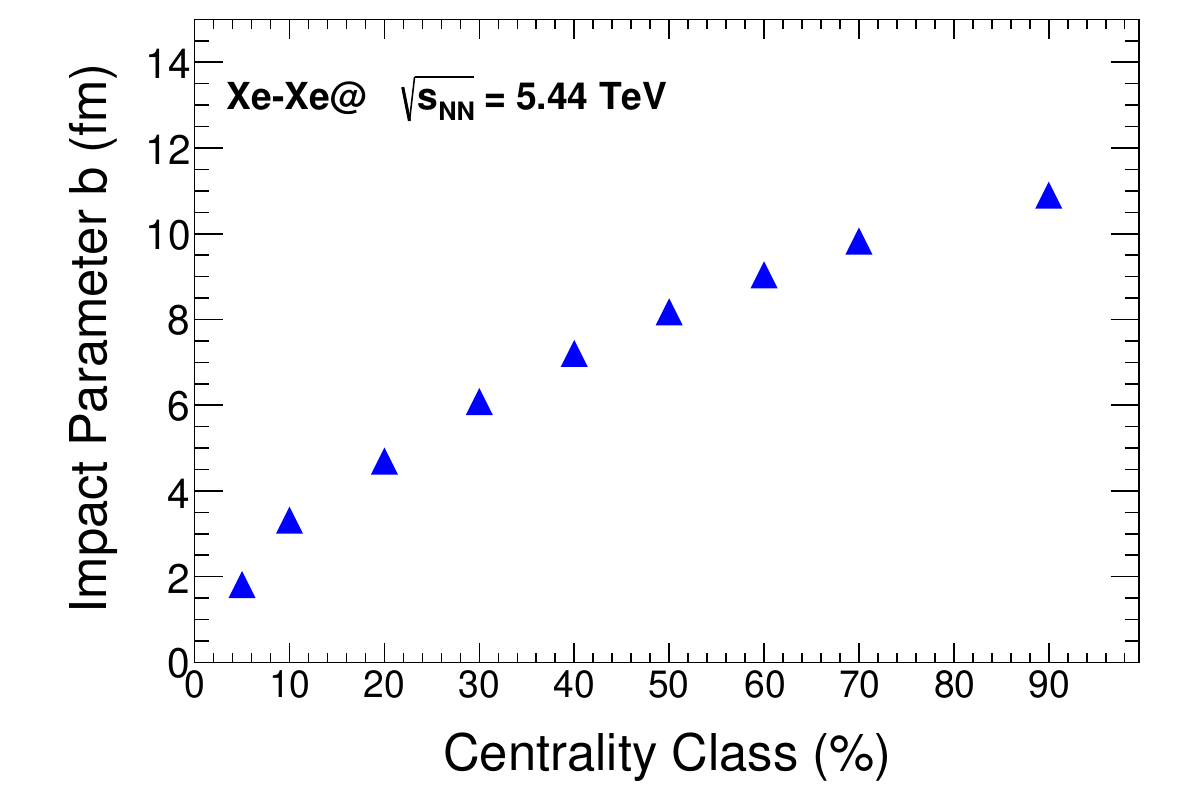}
	\caption{\label{5} Variation of impact parameter (b) with centrality in in Xe-Xe collisions at $\sqrt{s_{NN}} = 5.44$ TeV.}
\end{figure}

\begin{figure}[htb]
	\centering
	\includegraphics[width=0.91\linewidth]{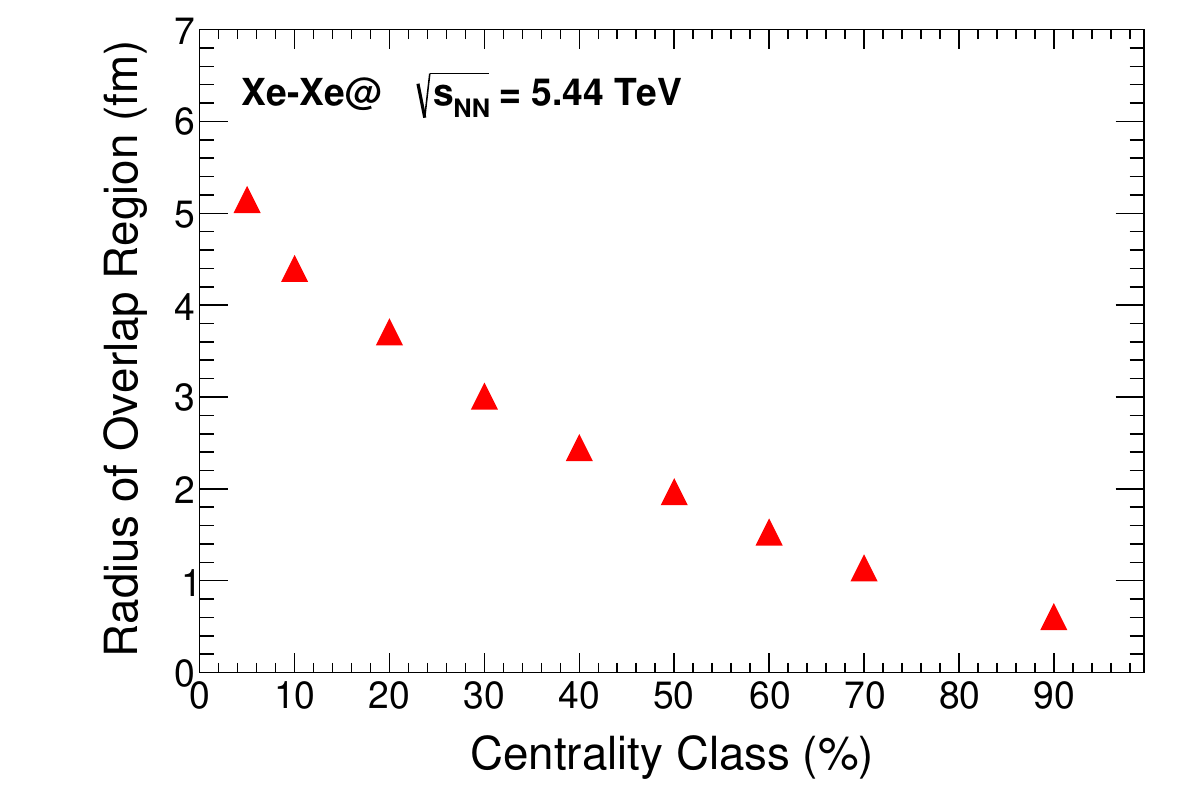}
	\caption{\label{6} Variation of radius of overlap region $(r_{overlap})$ with centrality in in Xe-Xe collisions at $\sqrt{s_{NN}} = 5.44$ TeV.}
\end{figure}

\begin{figure}[htb]
	\centering
	\includegraphics[width=0.95\linewidth]{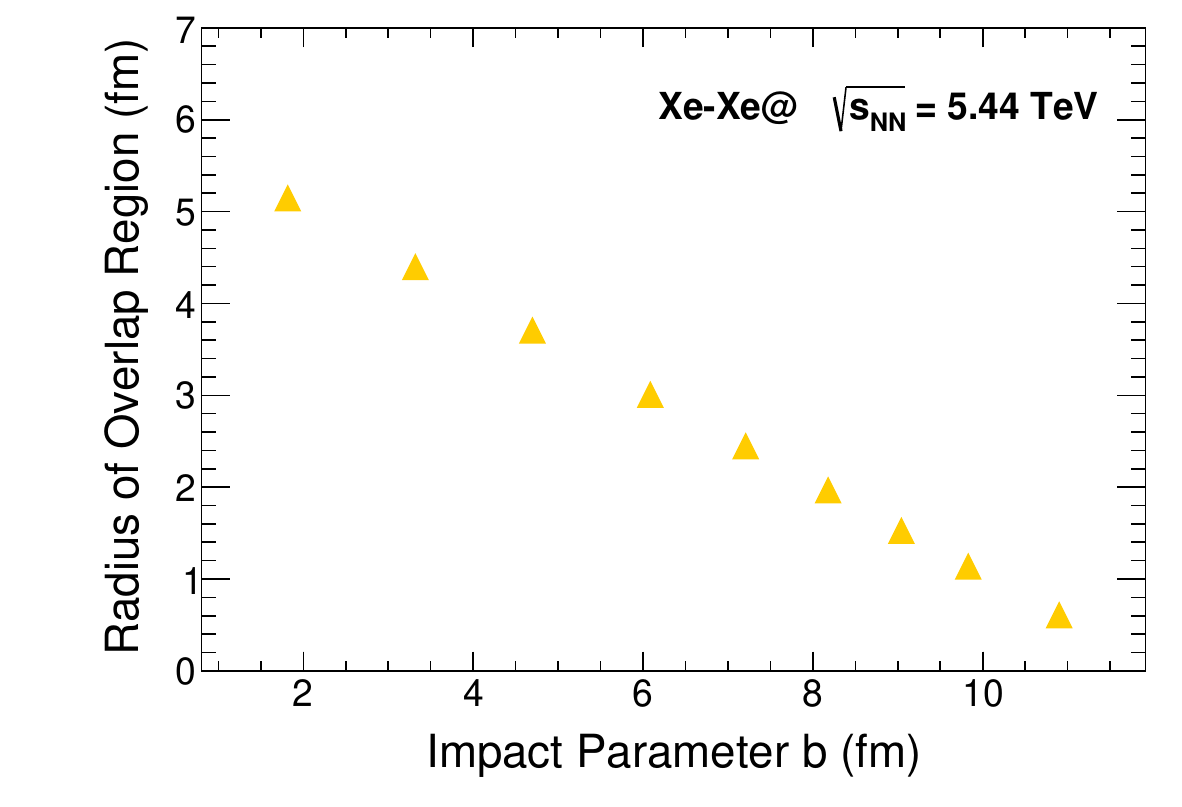}
	\caption{\label{7} Variation of radius of overlap region $(r_{overlap})$ with impact parameter (b) in in Xe-Xe collisions at $\sqrt{s_{NN}} = 5.44$ TeV.}
\end{figure}

We assumed that the colliding nuclei are of spherical shape~\cite{Gaudefroy:2009zza,Sarazin:1999pe,Jia:2021tzt,Choi:2022rdj,Zhou:2009sp}. We also assumed that the overlap region between the two nuclei colliding with each other is of circular shape. The radius of overlap region will depend upon the impact parameter (b), defined as the distance between the two centers of the nuclei colliding with one another. We know from the geometry of nuclear collisions that, when b = 0 then the radius of overlap region will be the radius of any of the two colliding nuclei. This is the case of top central collision. So for $b = 0, \; r_{overlap} = r_{Xe} = 6.06$ fm. The geometrical representation of two colliding nuclei is shown in figure~\ref{2}. When $b = r_{Xe}$, then $r_{overlap} = \frac{r_{Xe}}{2} = \frac{6.06}{2} = 3.03$ fm. This is the case of mid central collision. The geometrical representation of two colliding nuclei is shown in figure~\ref{3}. For $b = 2r_{Xe}$, then $r_{overlap} = 0$. This is the case of peripheral collision. The geometric view of two colliding nuclei is shown in figure~\ref{4}. 
\\
Based on the above observations, we have defined a relation between the radius of the overlap region ($r_{overlap}$), radius of Xenon nucleus ($r_{Xe}$) and impact parameter (b), such that the above conditions are satisfied.
\begin{equation}
	r_{overlap} = r_{Xe} - \frac{b}{2}
\end{equation}
Now using the impact parameter (b) values for different centrality classes in Xe-Xe collision at $\sqrt{s_{NN}} = 5.44$ TeV from the blast wave model~\cite{Lao:2021wub} and $r_{Xe}$ = 6.06 fm, we have calculated the radius of overlap region $(r_{overlap})$ for different centrality classes by using above defined eq. (3). The $r_{overlap}$ calculated values for different centrality classes are shown in table~\ref{table1}. The plot of impact parameter (b) versus centrality is shown in figure~\ref{5}. The plot shows that the impact parameter (b) increases with the centrality class.
\begin{table*}
\caption{\label{table1} Radius of overlap region $(r_{overlap})$ calculated by using impact parameter (b) values at $r_{Xe} = 6.06$ fm in eq. (3) for different centrality classes in Xe-Xe collisions at $\sqrt{s_{NN}} = 5.44$ TeV.}	
\begin{ruledtabular}
	\begin{tabular}{ccc}
		&\multicolumn{0}{c}{}\\
		Centrality class (\%)&Impact parameter (b) (fm)&$r_{overlap}$ (fm)\\ \hline
		0-5&1.817&5.15\\
		5-10&3.320&4.40\\
		10-20&4.698&3.71\\
		20-30&6.086&3.01\\
		30-40&7.209&2.45\\
		40-50&8.179&1.97\\
		50-60&9.042&1.53\\
		60-70&9.829&1.14\\
		70-90&10.900&0.61\\
	\end{tabular}
\end{ruledtabular}
\end{table*}
\\\\
The plot of radius of overlap region $(r_{overlap})$ versus centrality is shown in figure~\ref{6}. The plot shows that the radius of overlap region $(r_{overlap})$ decreases with the centrality class.
\\
The plot of radius of overlap region $(r_{overlap})$ versus impact parameter (b) is shown in figure~\ref{7}. The plot shows that the radius of overlap region $(r_{overlap})$ decreases with the impact parameter (b).
\subsection{Area of overlap region}
\begin{figure}[htb]
	\centering
	\includegraphics[width=1\linewidth]{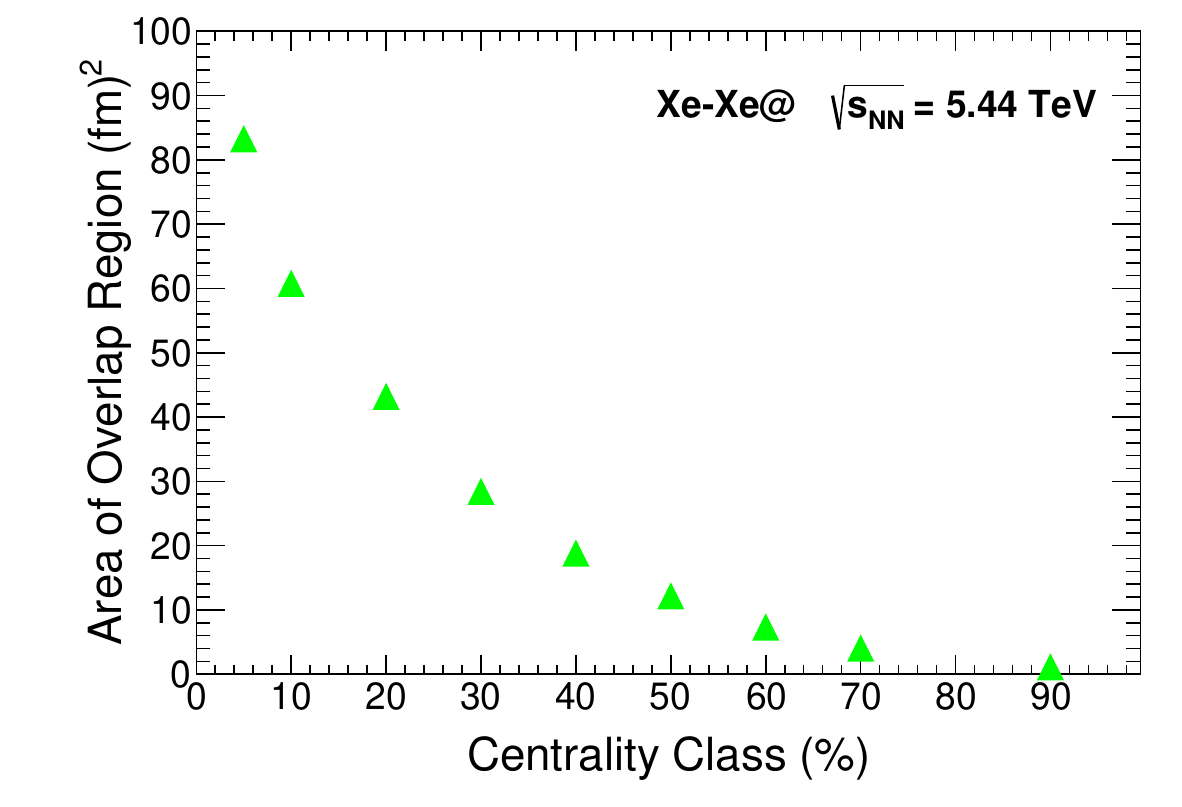}
	\caption{\label{8} Variation of area of overlap region $(A_{overlap})$ with centrality in Xe-Xe collisions at $\sqrt{s_{NN}} = 5.44$ TeV.}
\end{figure}

The area of overlap region $(A_{overlap})$ was calculated by using the formula $A_{overlap} = \pi r_{overlap}^{2}$~\cite{McLerran:2013oju} for different centrality classes in Xe-Xe collisions at $\sqrt{s_{NN}} = 5.44$ TeV. As discussed above, we have assumed that the colliding nuclei are of spherical shape and the overlap region after the collision between two colliding nuclei as of circular shape. The $A_{overlap}$ calculated values for different centrality classes are shown in table~\ref{table2}. The plot for area of overlap region $(A_{overlap})$ versus centrality is shown in figure~\ref{8}. The plot shows that the area of overlap region $(A_{overlap})$ decreases with the centrality class.
\begin{table*}
	\caption{\label{table2}Area of overlap region $(A_{overlap})$ calculated by using the radius of overlap region $(r_{overlap})$ as given in table~\ref{2} in the formula ($A_{overlap} = \pi r_{overlap}^{2}$) for different centrality classes in Xe-Xe collisions at $\sqrt{s_{NN}} = 5.44$ TeV.}\vspace{0cm}
	\begin{ruledtabular}
		\begin{tabular}{ccc}
			&\multicolumn{0}{c}{}\\
			Centrality class (\%)&$r_{overlap}$ (fm)&$A_{overlap}$ (fm)$^{2}$\\ \hline
			0-5&5.15&83.28\\
			5-10&4.40&60.79\\
			10-20&3.71&43.21\\
			20-30&3.01&28.44\\
			30-40&2.45&18.84\\
			40-50&1.97&12.18\\
			50-60&1.53&7.35\\
			60-70&1.14&4.08\\
			70-90&0.61&1.16\\
		\end{tabular}
	\end{ruledtabular}
\end{table*}
\subsection{Bjorken initial energy density}
The Bjorken initial energy density was calculated for different centrality classes for three different cases by using Bjorken formula as given in eq. (1), in Xe-Xe collisions at $\sqrt{s_{NN}} = 5.44$ TeV. The average transverse masss ($<m_{t}>$) of pions used is equal to 0.562 GeV. The charged particle rapidity density ($\frac{dN_{ch}}{dy}$) used is taken from ALICE~\cite{ALICE:2018cpu}. The formation time $\tau_{0}$ used is 1 $(fm/c)$ for all the centrality classes in Case I and II and for Case III, $\tau_{0}$ used varies from 1 to 2.6 $(fm/c)$, while going from central to the peripheral collisions. The Bjorken initial energy density ($\epsilon_{B}$) was then calculated for different centrality classes for three different cases as will be discussed in next section.
\section{\label{D} Results}\vspace{2cm}
\subsection{Bjorken initial energy density ($\epsilon_{B}$) values calculated for different centrality classes in Xe-Xe collisions at $\sqrt{s_{NN}} = 5.44$ TeV in Case I}
\begin{figure}[htb]
	\centering
	\includegraphics[width=1\linewidth]{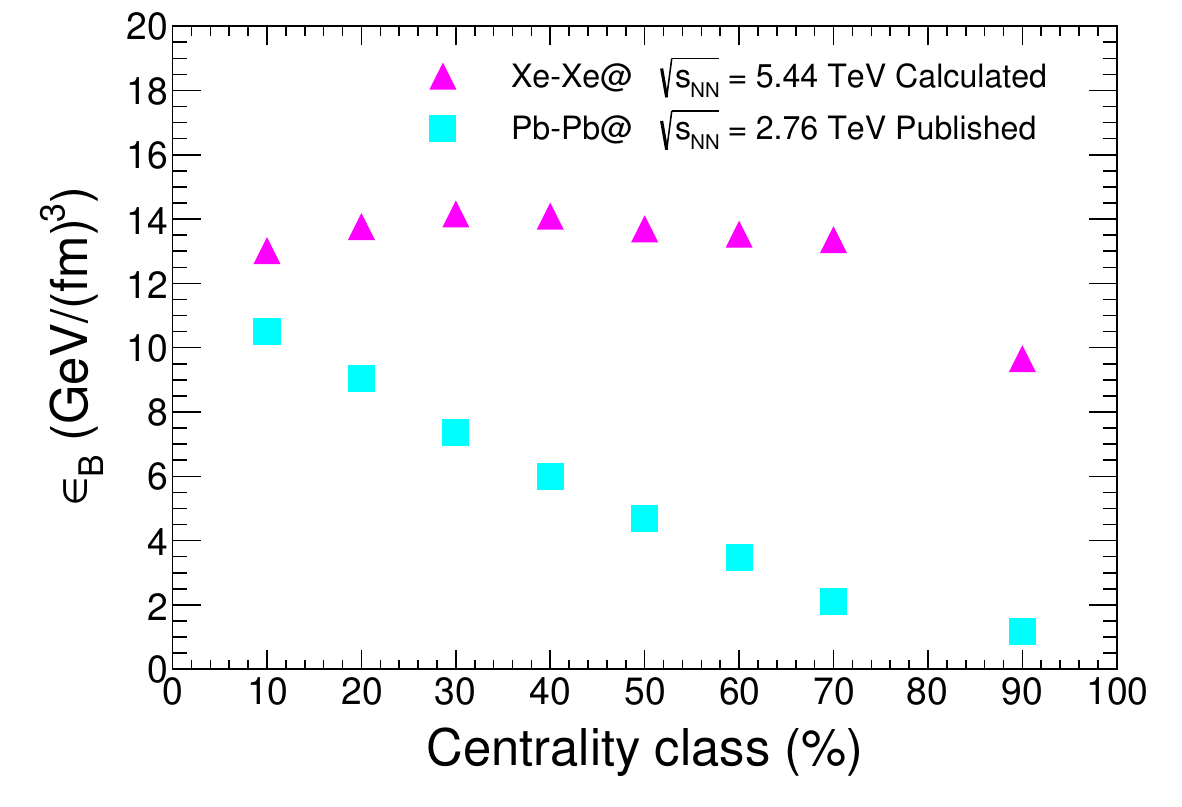}
	\includegraphics[width=1\linewidth]{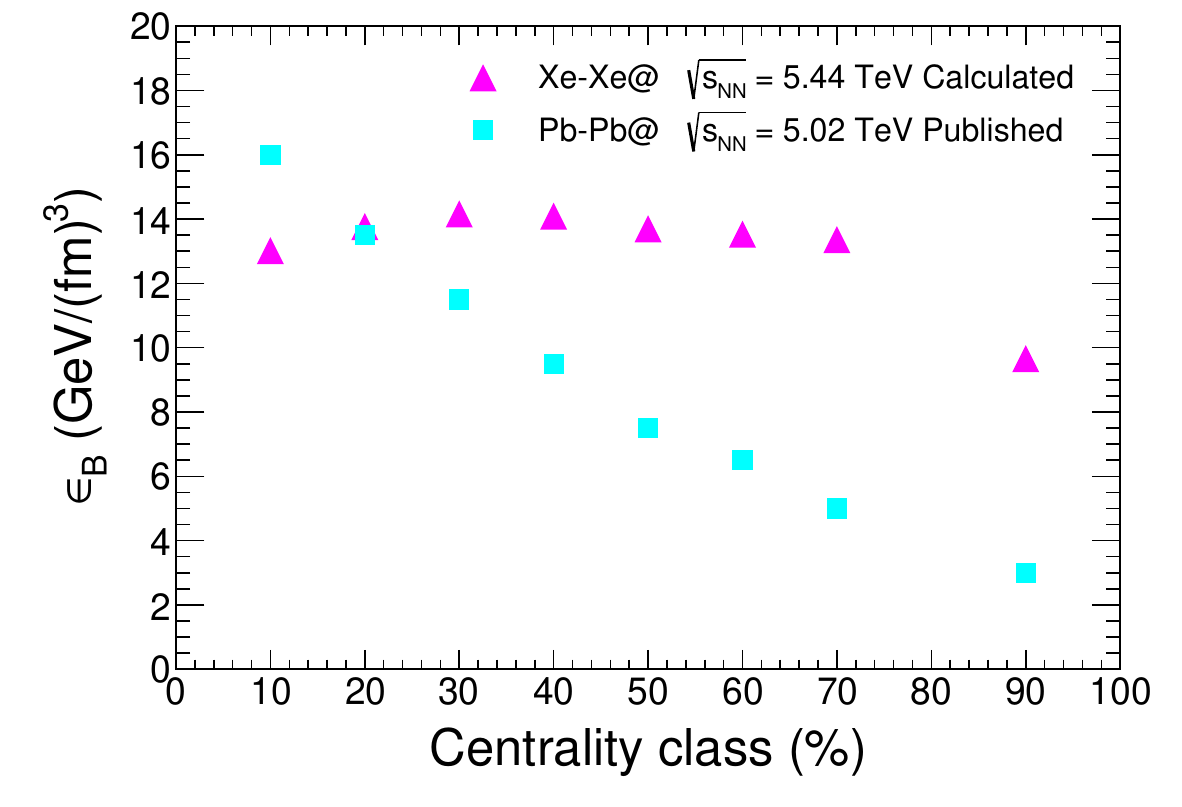}
	\caption{\label{9} Upper: Variation of Bjorken initial energy density ($\epsilon_{B}$) with centrality in Xe-Xe collisions at $\sqrt{s_{NN}} = 5.44$ TeV compared with Pb-Pb results at $\sqrt{s_{NN}} = 2.76$ TeV. Lower: Variation of Bjorken initial energy density ($\epsilon_{B}$) with centrality in Xe-Xe collisions at $\sqrt{s_{NN}} = 5.44$ TeV compared with Pb-Pb results at $\sqrt{s_{NN}} = 5.02$ TeV.}
\end{figure}

In Case I, we have fixed the formation time $\tau_{0}$ as 1 $(fm/c)$ for all the centrality classes and varied the area of overlap region $(A_{overlap})$ as given in table~\ref{table3} for different centrality classes. The Bjorken initial energy density ($\epsilon_{B}$) was then calculated for different centrality classes as are shown in table~\ref{table3}. The plot of Bjorken initial energy density ($\epsilon_{B}$) versus centrality is shown in figure~\ref{9}. Upper panel of the figure shows the variation of Bjorken initial energy density ($\epsilon_{B}$) with centrality compared with the Pb-Pb results at $\sqrt{s_{NN}} = 2.76$ TeV. In the centrality class (5-10)\%, $\epsilon_{B}$ in Xe-Xe collisions at $\sqrt{s_{NN}} = 5.44$ TeV is 19.35\% more than in Pb-Pb collisions at $\sqrt{s_{NN}} = 2.76$ TeV. This may be interpreted as, the system size in Xe-Xe collisions at $\sqrt{s_{NN}} = 5.44$ TeV is 19.35\% bigger than Pb-Pb collisions at $\sqrt{s_{NN}} = 2.76$ TeV. The lower panel of the figure shows the variation of Bjorken initial energy density ($\epsilon_{B}$) with centrality compared with the Pb-Pb results at $\sqrt{s_{NN}} = 5.02$ TeV. In the centrality class (5-10)\%, $\epsilon_{B}$ in Xe-Xe collisions at $\sqrt{s_{NN}} = 5.44$ TeV is 18.62\% less than in Pb-Pb collisions at $\sqrt{s_{NN}} = 5.02$ TeV. This may be interpreted as, the system size in Xe-Xe collisions at $\sqrt{s_{NN}} = 5.44$ TeV is 18.62\% smaller than Pb-Pb collisions at $\sqrt{s_{NN}} = 5.02$ TeV. Also the $\epsilon_{B}$ values in Xe-Xe collisions at $\sqrt{s_{NN}} = 5.44$ TeV initially increases from top to mid central collisions and then decreases from mid central to peripheral collisions. This may be due to the fixing of the formation time $\tau_{0}$ as 1 $(fm/c)$ for all the centrality classes, whereas the $\tau_{0}$ is increasing while going from the central to the peripheral collisions.
\begin{table*}
	\caption{\label{table3} Bjorken initial energy density $(\epsilon_{B})$ calculated by using the charged particle rapidity density ($\frac{dN_{ch}}{dy}$)~\cite{ALICE:2018cpu}, area of overlap region $(A_{overlap})$ and the formation time $\tau_{0}$ as given in this table in the eq. (1) for different centrality classes in Xe-Xe collisions at $\sqrt{s_{NN}} = 5.44$ TeV.}
	\begin{ruledtabular}
		\begin{tabular}{ccccc}
			&\multicolumn{0}{c}{}\\
			Centrality class (\%)&$\frac{dN_{ch}}{d\eta}$&$A_{overlap}$ (fm)$^{2}$&$\tau_{0}$ (fm/c)&$\epsilon_{B}$ ($GeV/(fm)^{3}$)\\ \hline
			0-5&1167&83.28&1&11.81\\
			5-10&939&60.79&1&13.02\\
			10-20&706&43.21&1&13.77\\
			20-30&478&28.44&1&14.16\\
			30-40&315&18.84&1&14.09\\
			40-50&198&12.18&1&13.70\\
			50-60&118&7.35&1&13.53\\
			60-70&64.7&4.08&1&13.36\\
			70-90&13.3&1.16&1&9.66\\
		\end{tabular}
	\end{ruledtabular}
\end{table*}\\\\\\\\\\\
\\\\\\
\subsection{Bjorken initial energy density ($\epsilon_{B}$) values calculated for different centrality classes in Xe-Xe collisions at $\sqrt{s_{NN}} = 5.44$ TeV in Case II}
\begin{figure}[htb]
	\centering
	\includegraphics[width=0.98\linewidth]{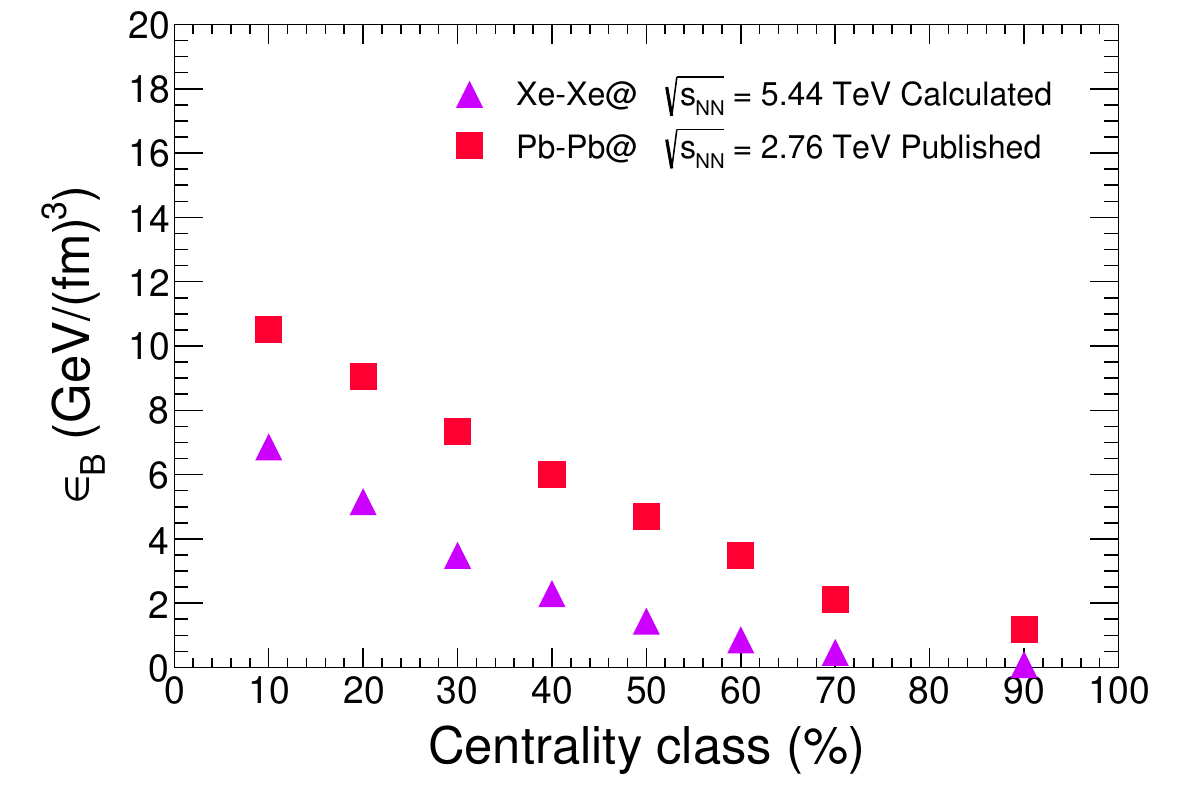}\par
	\includegraphics[width=0.98\linewidth]{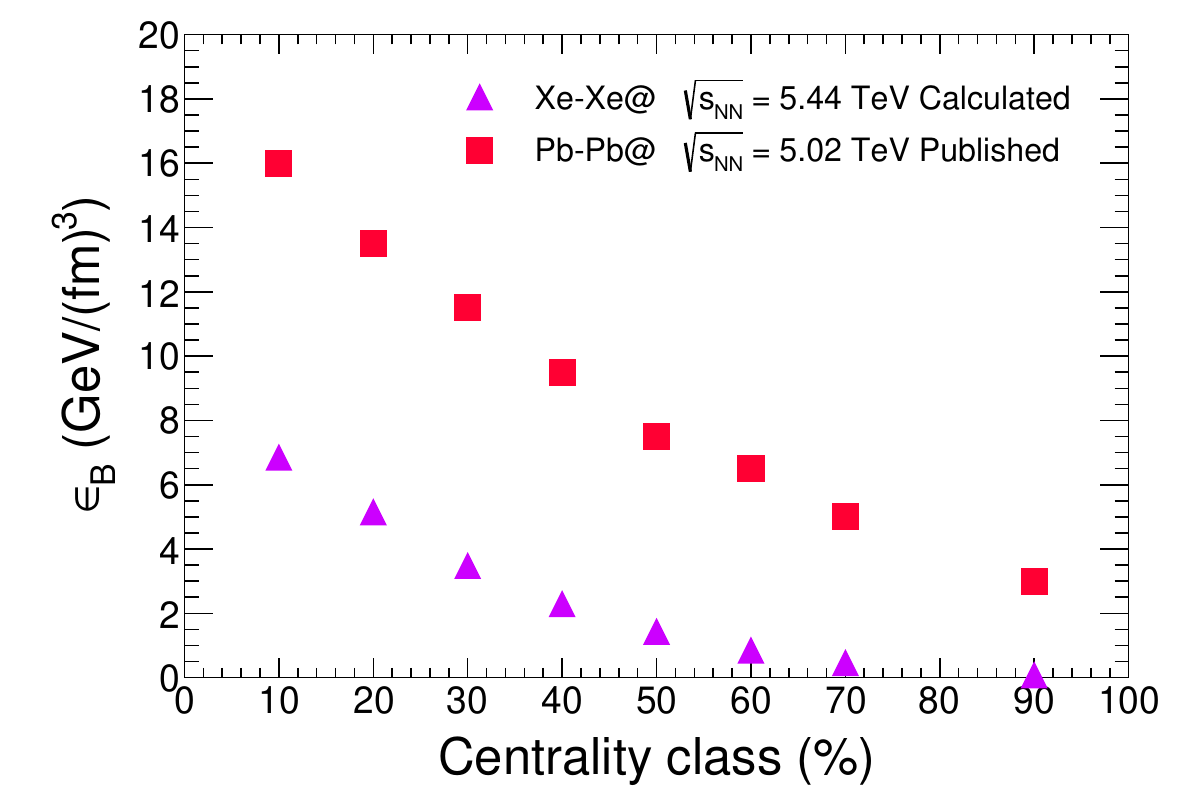}
	\caption{\label{10} Upper: Variation of Bjorken initial energy density ($\epsilon_{B}$) with centrality in Xe-Xe collisions at $\sqrt{s_{NN}} = 5.44$ TeV compared with Pb-Pb results at $\sqrt{s_{NN}} = 2.76$ TeV. Lower: Variation of Bjorken initial energy density ($\epsilon_{B}$) with centrality in Xe-Xe collisions at $\sqrt{s_{NN}} = 5.44$ TeV compared with Pb-Pb results at $\sqrt{s_{NN}} = 5.02$ TeV.}
\end{figure}

In Case II, we have fixed the formation time $\tau_{0}$ as 1 $(fm/c)$ for all the centrality classes as given in table~\ref{4} and also fixed the area of overlap region $(A_{overlap})$ as 115.30 (fm)$^{2}$, calculated by using $r_{overlap} = r_{Xe} = $ 6.06 fm and area of overlap region $(A_{overlap}) = \pi r_{overlap}^{2} = $ 115.30 (fm)$^{2}$ as given in table~\ref{table4}. The Bjorken initial energy density ($\epsilon_{B}$) was then calculated for different centrality classes as given in table~\ref{table4}. The plot of Bjorken initial energy density ($\epsilon_{B}$) versus centrality is shown in figure~\ref{10}. Upper panel of the figure shows the variation of Bjorken initial energy density ($\epsilon_{B}$) with centrality compared with the Pb-Pb results at $\sqrt{s_{NN}} = 2.76$ TeV. In the centrality class (5-10)\%, $\epsilon_{B}$ in Xe-Xe collisions at $\sqrt{s_{NN}} = 5.44$ TeV is 34.66\% less than in Pb-Pb collisions at $\sqrt{s_{NN}} = 2.76$ TeV. This may be interpreted as, the system size in Xe-Xe collisions at $\sqrt{s_{NN}} = 5.44$ TeV is 34.66\% smaller than Pb-Pb collisions at $\sqrt{s_{NN}} = 2.76$ TeV. Lower panel of the figure shows the variation of Bjorken initial energy density ($\epsilon_{B}$) with centrality compared with the Pb-Pb results at $\sqrt{s_{NN}} = 5.02$ TeV. In the centrality class (5-10)\%, $\epsilon_{B}$ in Xe-Xe collisions at $\sqrt{s_{NN}} = 5.44$ TeV is 57.12\% less than in Pb-Pb collisions at $\sqrt{s_{NN}} = 5.02$ TeV. This may be interpreted as, the system size in Xe-Xe collisions at $\sqrt{s_{NN}} = 5.44$ TeV is 57.12\% smaller than Pb-Pb collisions at $\sqrt{s_{NN}} = 5.02$ TeV. Also the $\epsilon_{B}$ values in Xe-Xe collisions at $\sqrt{s_{NN}} = 5.44$ TeV decreases from top central to peripheral collisions, which is expected. The fixing of the formation time $\tau_{0}$ as 1 $(fm/c)$ compensates the fixing of the area of overlap region as $(A_{overlap})$ as 115.30 (fm)$^{2}$ for all the centrality classes and thus gives the expected decrease of the $\epsilon_{B}$ values.
\begin{table*}
	\caption{\label{table4} Bjorken initial energy density $(\epsilon_{B})$ calculated by using the charged particle rapidity density ($\frac{dN_{ch}}{dy}$)~\cite{ALICE:2018cpu}, area of overlap region $(A_{overlap})$, the formation time $\tau_{0}$ as given in this table in the eq. (1) for different centrality classes in Xe-Xe collisions at $\sqrt{s_{NN}} = 5.44$ TeV.}
	\begin{ruledtabular}
		\begin{tabular}{ccccc}
			&\multicolumn{0}{c}{}\\
			Centrality class (\%)&$\frac{dN_{ch}}{d\eta}$&$A_{overlap}$ (fm)$^{2}$&$\tau_{0}$ (fm/c)&$\epsilon_{B}$ ($GeV/(fm)^{3}$)\\ \hline
			0-5&1167&115.30&1&8.53\\
			5-10&939&115.30&1&6.86\\
			10-20&706&115.30&1&5.16\\
			20-30&478&115.30&1&3.49\\
			30-40&315&115.30&1&2.30\\
			40-50&198&115.30&1&1.44\\
			50-60&118&115.30&1&0.86\\
			60-70&64.7&115.30&1&0.47\\
			70-90&13.3&115.30&1&0.09\\
		\end{tabular}
	\end{ruledtabular}
\end{table*}\\\\\\\\\\
\\\\\\
\subsection{Bjorken initial energy density ($\epsilon_{B}$) values calculated for different centrality classes in Xe-Xe collisions at $\sqrt{s_{NN}} = 5.44$ TeV in Case III}
\begin{figure}[htb]
	\centering
	\includegraphics[width=0.98\linewidth]{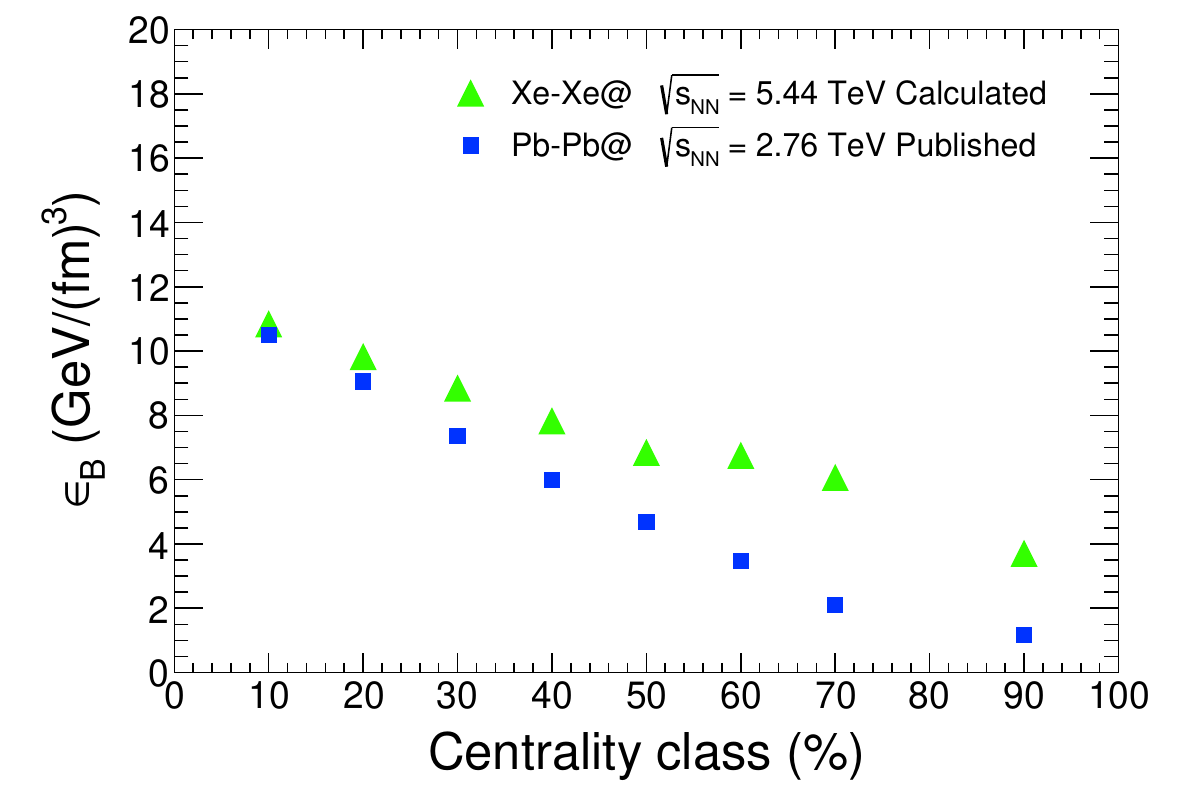}\par
	\includegraphics[width=0.98\linewidth]{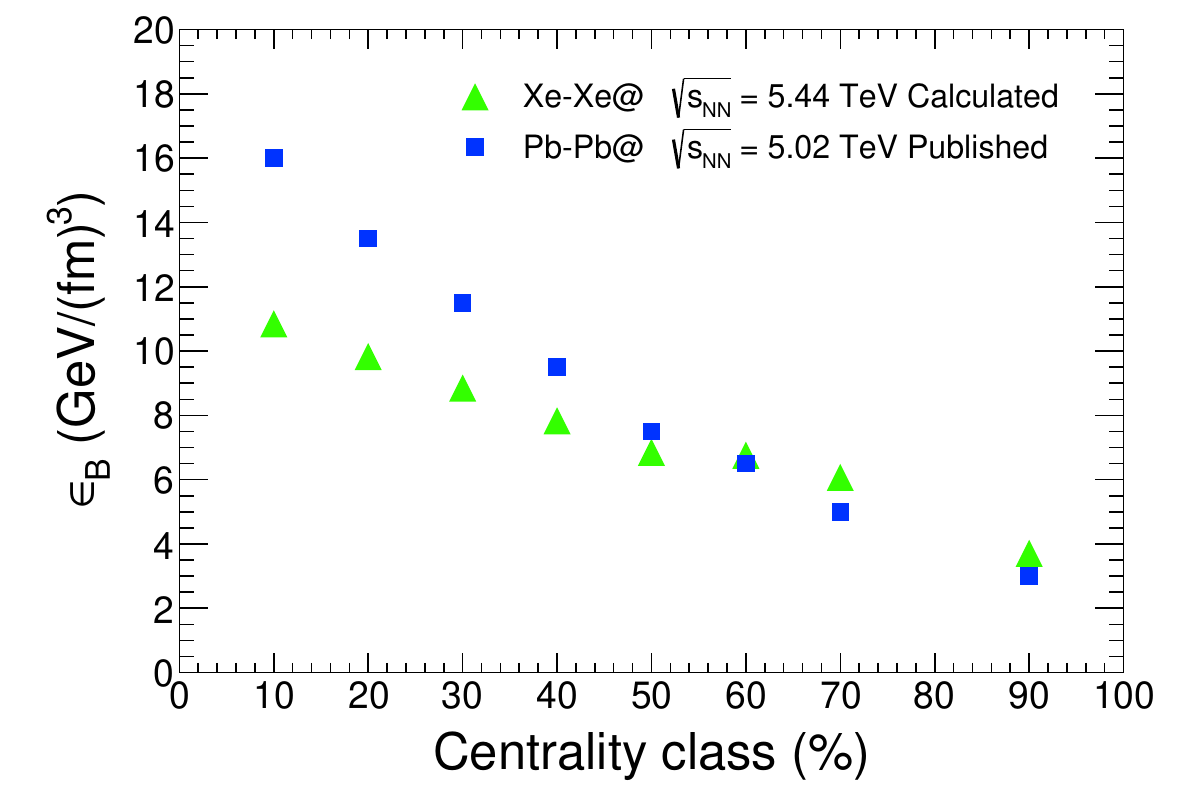}
	\caption{\label{11} Upper: Variation of Bjorken initial energy density ($\epsilon_{B}$) with centrality in Xe-Xe collisions at $\sqrt{s_{NN}} = 5.44$ TeV compared with Pb-Pb results at $\sqrt{s_{NN}} = 2.76$ TeV. Lower: Variation of Bjorken initial energy density ($\epsilon_{B}$) with centrality in Xe-Xe collisions at $\sqrt{s_{NN}} = 5.44$ TeV compared with Pb-Pb results at $\sqrt{s_{NN}} = 5.02$ TeV.}
\end{figure}

In case III, we have varied the formation time $\tau_{0}$ from 1 to 2.6 $(fm/c)$~\cite{Singh:2021evv} from top central to peripheral collisions as given in table~\ref{table5} and also varied the area of overlap region $(A_{overlap})$, which is calculated by using $r_{overlap}$ from eq. (3) and $A_{overlap} = \pi r_{overlap}^{2}$ for different centrality classes as given in table~\ref{table5}. The Bjorken initial energy density ($\epsilon_{B}$) was then calculated for different centrality classes as given in table~\ref{table5}. The plot of Bjorken initial energy density ($\epsilon_{B}$) versus centrality is shown in figure~\ref{11}. Upper panel of the figure shows the variation of Bjorken initial energy density ($\epsilon_{B}$) with centrality compared with the Pb-Pb results at $\sqrt{s_{NN}} = 2.76$ TeV. In the centrality class (5-10)\%, $\epsilon_{B}$ in Xe-Xe collisions at $\sqrt{s_{NN}} = 5.44$ TeV is 3.22\% more than in Pb-Pb collisions at $\sqrt{s_{NN}} = 2.76$ TeV. This may be interpreted as, the system size in Xe-Xe collisions at $\sqrt{s_{NN}} = 5.44$ TeV is 3.22\% bigger than Pb-Pb collisions at $\sqrt{s_{NN}} = 2.76$ TeV. Lower panel of the figure shows the variation of Bjorken initial energy density ($\epsilon_{B}$) with centrality compared with the Pb-Pb results at $\sqrt{s_{NN}} = 5.02$ TeV. In the centrality class (5-10)\%, $\epsilon_{B}$ in Xe-Xe collisions at $\sqrt{s_{NN}} = 5.44$ TeV is 32.18\% less than in Pb-Pb collisions at $\sqrt{s_{NN}} = 5.02$ TeV. This may be interpreted as, the system size in Xe-Xe collisions at $\sqrt{s_{NN}} = 5.44$ TeV is 32.18\% smaller than Pb-Pb collisions at $\sqrt{s_{NN}} = 5.02$ TeV. Also the $\epsilon_{B}$ values in Xe-Xe collisions at $\sqrt{s_{NN}} = 5.44$ TeV decreases from top central to peripheral collisions, which is expected. The increase in the formation time $\tau_{0}$ and the decrease in the area of overlap region $(A_{overlap})$ from top central to peripheral collisions gives the expected decrease of the $\epsilon_{B}$ values.
\begin{table*}
 \caption{\label{table5} Bjorken initial energy density $(\epsilon_{B})$ calculated by using the charged particle rapidity density ($\frac{dN_{ch}}{dy}$)~\cite{ALICE:2018cpu}, area of overlap region $(A_{overlap})$, the formation time $\tau_{0}$ as given in this table in the eq. (1) for different centrality classes in Xe-Xe collisions at $\sqrt{s_{NN}} = 5.44$ TeV.}
	\begin{ruledtabular}
		\begin{tabular}{ccccc}
			&\multicolumn{0}{c}{}\\
			Centrality class (\%)&$\frac{dN_{ch}}{d\eta}$&$A_{overlap}$ (fm)$^{2}$&$\tau_{0}$ (fm/c)&$\epsilon_{B}$ ($GeV/(fm)^{3}$)\\ \hline
			0-5&1167&83.28&1&11.81\\
			5-10&939&60.79&1.2&10.85\\
			10-20&706&43.21&1.4&9.83\\
			20-30&478&28.44&1.6&8.85\\
			30-40&315&18.84&1.8&7.83\\
			40-50&198&12.18&2.0&6.85\\
			50-60&118&7.35&2.2&6.76\\
			60-70&64.7&4.08&2.4&6.07\\
			70-90&13.3&1.16&2.6&3.71\\
		\end{tabular}
	\end{ruledtabular}
\end{table*}\\\\\\\
\section{\label{E} Summary}
In this work, we have calculated the Bjorken initial energy density ($\epsilon_{B}$) in different centrality classes in Xe-Xe collisions at $\sqrt{s_{NN}} = 5.44$ TeV for three different cases.\par 
In Case I, we have calculated the $\epsilon_{B}$ values for different centrality classes by fixing the formation time $\tau_{0}$ as 1 $(fm/c)$ and varied the area of overlap region calculated by using $A_{overlap} = \pi r_{overlap}^{2}$, where $r_{overlap}$ is obtained from eq. (3). It is observed that the $\epsilon_{B}$ values first increase while going from top central to mid central and then decrease while going from mid central to peripheral collisions, which is not expected and may be due to the fixing of the $\tau_{0}$. In the centrality class (5-10)\%, $\epsilon_{B}$ in Xe-Xe collisions at $\sqrt{s_{NN}} = 5.44$ TeV is 19.35\% more than in Pb-Pb collisions at $\sqrt{s_{NN}} = 2.76$ TeV and 18.62\% less than in Pb-Pb collisions at $\sqrt{s_{NN}} = 5.02$ TeV. This may be interpreted as, the system size in Xe-Xe collisions at $\sqrt{s_{NN}} = 5.44$ TeV is 19.35\% bigger than Pb-Pb collisions at $\sqrt{s_{NN}} = 2.76$ TeV and 18.62\% smaller than Pb-Pb collisions at $\sqrt{s_{NN}} = 5.02$ TeV.\par
In Case II, we have calculated the $\epsilon_{B}$ values for different centrality classes by fixing both the formation time ($\tau_{0}$) as 1 $(fm/c)$ and the area of overlap region as 115.30 $(fm)^{2}$ calculated by $A_{overlap} = \pi r_{overlap}^{2}$, where $r_{overlap} = r_{Xe} = 6.06$ fm. It is observed that the $\epsilon_{B}$ values decrease while going from top central to peripheral collisions, which is expected. Thus fixing the $A_{overlap}$ for all centrality classes is compensated by fixing the $\tau_{0}$. In the centrality class (5-10)\%, $\epsilon_{B}$ in Xe-Xe collisions at $\sqrt{s_{NN}} = 5.44$ TeV is 34.66\% less than in Pb-Pb collisions at $\sqrt{s_{NN}} = 2.76$ TeV and 57.12\% less than in Pb-Pb collisions at $\sqrt{s_{NN}} = 5.02$ TeV. This may be interpreted as, the system size in Xe-Xe collisions at $\sqrt{s_{NN}} = 5.44$ TeV is 34.66\% smaller than Pb-Pb collisions at $\sqrt{s_{NN}} = 2.76$ TeV and 57.12\% smaller than Pb-Pb collisions at $\sqrt{s_{NN}} = 5.02$ TeV.\par 
In Case III, we have calculated the $\epsilon_{B}$ values for different centrality classes by varying both the formation time ($\tau_{0}$) and the area of overlap region $(A_{overlap})$ calculated by $A_{overlap} = \pi r_{overlap}^{2}$, where $r_{overlap}$ is obtained from eq. (3). It is observed that the $\epsilon_{B}$ values decrease while going from top central to peripheral collisions, which is expected. In the centrality class (5-10)\%, $\epsilon_{B}$ in Xe-Xe collisions at $\sqrt{s_{NN}} = 5.44$ TeV is 3.22\% more than in Pb-Pb collisions at $\sqrt{s_{NN}} = 2.76$ TeV and 32.18\% less than in Pb-Pb collisions at $\sqrt{s_{NN}} = 5.02$ TeV. This may be interpreted as, the system size in Xe-Xe collisions at $\sqrt{s_{NN}} = 5.44$ TeV is 3.22\% bigger than Pb-Pb collisions at $\sqrt{s_{NN}} = 2.76$ TeV and 32.18\% smaller than Pb-Pb collisions at $\sqrt{s_{NN}} = 5.02$ TeV. In the nutshell, out of three cases Case II and Case III are giving the expected results. The $\epsilon_{B}$ value for top central and mid central collisions in Case II and for all centrality classes in Case III is greater than 1 $GeV/(fm)^{3}$. This indicates the possibility of QGP medium formation in top central and mid central collisions in Case II and both central and peripheral collisions in Case III. The results obtained in the Case III seems more suitable than Case II.
\nocite{*}

\bibliography{apssamp_V1}

\end{document}